\documentclass[aps,prl,twocolumn,superscriptaddress,showpacs,letterpaper]{revtex4}

\usepackage{graphicx}
\usepackage{amsmath, amsthm, amssymb}

\usepackage{xspace}
\usepackage{color}

\newcommand{\Jpsi}{\ensuremath{J\!/\!\psi}\xspace}
\newcommand{\X}{\ensuremath{X(3872)}\xspace}

\begin{document}

\title{Precision Measurement of the \X Mass in $\Jpsi\pi^+\pi^-$ Decays}

\affiliation{Institute of Physics, Academia Sinica, Taipei, Taiwan 11529, Republic of China} 
\affiliation{Argonne National Laboratory, Argonne, Illinois 60439} 
\affiliation{University of Athens, 157 71 Athens, Greece} 
\affiliation{Institut de Fisica d'Altes Energies, Universitat Autonoma de Barcelona, E-08193, Bellaterra (Barcelona), Spain} 
\affiliation{Baylor University, Waco, Texas  76798} 
\affiliation{Istituto Nazionale di Fisica Nucleare Bologna, $^y$University of Bologna, I-40127 Bologna, Italy} 
\affiliation{Brandeis University, Waltham, Massachusetts 02254} 
\affiliation{University of California, Davis, Davis, California  95616} 
\affiliation{University of California, Los Angeles, Los Angeles, California  90024} 
\affiliation{University of California, San Diego, La Jolla, California  92093} 
\affiliation{University of California, Santa Barbara, Santa Barbara, California 93106} 
\affiliation{Instituto de Fisica de Cantabria, CSIC-University of Cantabria, 39005 Santander, Spain} 
\affiliation{Carnegie Mellon University, Pittsburgh, PA  15213} 
\affiliation{Enrico Fermi Institute, University of Chicago, Chicago, Illinois 60637}
\affiliation{Comenius University, 842 48 Bratislava, Slovakia; Institute of Experimental Physics, 040 01 Kosice, Slovakia} 
\affiliation{Joint Institute for Nuclear Research, RU-141980 Dubna, Russia} 
\affiliation{Duke University, Durham, North Carolina  27708} 
\affiliation{Fermi National Accelerator Laboratory, Batavia, Illinois 60510} 
\affiliation{University of Florida, Gainesville, Florida  32611} 
\affiliation{Laboratori Nazionali di Frascati, Istituto Nazionale di Fisica Nucleare, I-00044 Frascati, Italy} 
\affiliation{University of Geneva, CH-1211 Geneva 4, Switzerland} 
\affiliation{Glasgow University, Glasgow G12 8QQ, United Kingdom} 
\affiliation{Harvard University, Cambridge, Massachusetts 02138} 
\affiliation{Division of High Energy Physics, Department of Physics, University of Helsinki and Helsinki Institute of Physics, FIN-00014, Helsinki, Finland} 
\affiliation{University of Illinois, Urbana, Illinois 61801} 
\affiliation{The Johns Hopkins University, Baltimore, Maryland 21218} 
\affiliation{Institut f\"{u}r Experimentelle Kernphysik, Universit\"{a}t Karlsruhe, 76128 Karlsruhe, Germany} 
\affiliation{Center for High Energy Physics: Kyungpook National University, Daegu 702-701, Korea; Seoul National University, Seoul 151-742, Korea; Sungkyunkwan University, Suwon 440-746, Korea; Korea Institute of Science and Technology Information, Daejeon 305-806, Korea; Chonnam National University, Gwangju 500-757, Korea; Chonbuk National University, Jeonju 561-756, Korea} 
\affiliation{Ernest Orlando Lawrence Berkeley National Laboratory, Berkeley, California 94720} 
\affiliation{University of Liverpool, Liverpool L69 7ZE, United Kingdom} 
\affiliation{University College London, London WC1E 6BT, United Kingdom} 
\affiliation{Centro de Investigaciones Energeticas Medioambientales y Tecnologicas, E-28040 Madrid, Spain} 
\affiliation{Massachusetts Institute of Technology, Cambridge, Massachusetts  02139} 
\affiliation{Institute of Particle Physics: McGill University, Montr\'{e}al, Qu\'{e}bec, Canada H3A~2T8; Simon Fraser University, Burnaby, British Columbia, Canada V5A~1S6; University of Toronto, Toronto, Ontario, Canada M5S~1A7; and TRIUMF, Vancouver, British Columbia, Canada V6T~2A3} 
\affiliation{University of Michigan, Ann Arbor, Michigan 48109} 
\affiliation{Michigan State University, East Lansing, Michigan  48824}
\affiliation{Institution for Theoretical and Experimental Physics, ITEP, Moscow 117259, Russia} 
\affiliation{University of New Mexico, Albuquerque, New Mexico 87131} 
\affiliation{Northwestern University, Evanston, Illinois  60208} 
\affiliation{The Ohio State University, Columbus, Ohio  43210} 
\affiliation{Okayama University, Okayama 700-8530, Japan} 
\affiliation{Osaka City University, Osaka 588, Japan} 
\affiliation{University of Oxford, Oxford OX1 3RH, United Kingdom} 
\affiliation{Istituto Nazionale di Fisica Nucleare, Sezione di Padova-Trento, $^z$University of Padova, I-35131 Padova, Italy} 
\affiliation{LPNHE, Universite Pierre et Marie Curie/IN2P3-CNRS, UMR7585, Paris, F-75252 France} 
\affiliation{University of Pennsylvania, Philadelphia, Pennsylvania 19104}
\affiliation{Istituto Nazionale di Fisica Nucleare Pisa, $^{aa}$University of Pisa, $^{bb}$University of Siena and $^{cc}$Scuola Normale Superiore, I-56127 Pisa, Italy} 
\affiliation{University of Pittsburgh, Pittsburgh, Pennsylvania 15260} 
\affiliation{Purdue University, West Lafayette, Indiana 47907} 
\affiliation{University of Rochester, Rochester, New York 14627} 
\affiliation{The Rockefeller University, New York, New York 10021} 
\affiliation{Istituto Nazionale di Fisica Nucleare, Sezione di Roma 1, $^{dd}$Sapienza Universit\`{a} di Roma, I-00185 Roma, Italy} 

\affiliation{Rutgers University, Piscataway, New Jersey 08855} 
\affiliation{Texas A\&M University, College Station, Texas 77843} 
\affiliation{Istituto Nazionale di Fisica Nucleare Trieste/Udine, I-34100 Trieste, $^{ee}$University of Trieste/Udine, I-33100 Udine, Italy} 
\affiliation{University of Tsukuba, Tsukuba, Ibaraki 305, Japan} 
\affiliation{Tufts University, Medford, Massachusetts 02155} 
\affiliation{Waseda University, Tokyo 169, Japan} 
\affiliation{Wayne State University, Detroit, Michigan  48201} 
\affiliation{University of Wisconsin, Madison, Wisconsin 53706} 
\affiliation{Yale University, New Haven, Connecticut 06520} 
\author{T.~Aaltonen}
\affiliation{Division of High Energy Physics, Department of Physics, University of Helsinki and Helsinki Institute of Physics, FIN-00014, Helsinki, Finland}
\author{J.~Adelman}
\affiliation{Enrico Fermi Institute, University of Chicago, Chicago, Illinois 60637}
\author{T.~Akimoto}
\affiliation{University of Tsukuba, Tsukuba, Ibaraki 305, Japan}
\author{B.~\'{A}lvarez~Gonz\'{a}lez$^t$}
\affiliation{Instituto de Fisica de Cantabria, CSIC-University of Cantabria, 39005 Santander, Spain}
\author{S.~Amerio$^z$}
\affiliation{Istituto Nazionale di Fisica Nucleare, Sezione di Padova-Trento, $^z$University of Padova, I-35131 Padova, Italy} 

\author{D.~Amidei}
\affiliation{University of Michigan, Ann Arbor, Michigan 48109}
\author{A.~Anastassov}
\affiliation{Northwestern University, Evanston, Illinois  60208}
\author{A.~Annovi}
\affiliation{Laboratori Nazionali di Frascati, Istituto Nazionale di Fisica Nucleare, I-00044 Frascati, Italy}
\author{J.~Antos}
\affiliation{Comenius University, 842 48 Bratislava, Slovakia; Institute of Experimental Physics, 040 01 Kosice, Slovakia}
\author{G.~Apollinari}
\affiliation{Fermi National Accelerator Laboratory, Batavia, Illinois 60510}
% ...

\collaboration{CDF Collaboration\footnote{With visitors from $^a$University of Massachusetts Amherst, Amherst, Massachusetts 01003,
$^b$Universiteit Antwerpen, B-2610 Antwerp, Belgium, 
$^c$University of Bristol, Bristol BS8 1TL, United Kingdom,
$^d$Chinese Academy of Sciences, Beijing 100864, China, 
$^e$Istituto Nazionale di Fisica Nucleare, Sezione di Cagliari, 09042 Monserrato (Cagliari), Italy,
$^f$University of California Irvine, Irvine, CA  92697, 
$^g$University of California Santa Cruz, Santa Cruz, CA  95064, 
$^h$Cornell University, Ithaca, NY  14853, 
$^i$University of Cyprus, Nicosia CY-1678, Cyprus, 
$^j$University College Dublin, Dublin 4, Ireland,
$^k$University of Edinburgh, Edinburgh EH9 3JZ, United Kingdom, 
$^l$University of Fukui, Fukui City, Fukui Prefecture, Japan 910-0017
$^m$Kinki University, Higashi-Osaka City, Japan 577-8502
$^n$Universidad Iberoamericana, Mexico D.F., Mexico,
$^o$University of Iowa, Iowa City, IA  52242,
$^p$Queen Mary, University of London, London, E1 4NS, England,
$^q$University of Manchester, Manchester M13 9PL, England, 
$^r$Nagasaki Institute of Applied Science, Nagasaki, Japan, 
$^s$University of Notre Dame, Notre Dame, IN 46556,
$^t$University de Oviedo, E-33007 Oviedo, Spain, 
$^u$Texas Tech University, Lubbock, TX  79609, 
$^v$IFIC(CSIC-Universitat de Valencia), 46071 Valencia, Spain,
$^w$University of Virginia, Charlottesville, VA  22904,
$^x$Bergische Universit\"at Wuppertal, 42097 Wuppertal, Germany,
$^{ff}$On leave from J.~Stefan Institute, Ljubljana, Slovenia, 
}}
\noaffiliation
\pacs{14.40.Gx, 12.39.Mk, 13.25.Gv}

\begin{abstract}
  We present an analysis of the mass of the \X{} reconstructed via
its decay to $\Jpsi\pi^+\pi^-$ using $2.4$ fb$^{-1}$
of integrated luminosity from $p\bar{p}$ collisions at $\sqrt{s} = 1.96$ TeV, collected
with the CDF~II detector at the Fermilab Tevatron.
The possible existence of two nearby mass states is investigated.
Within the limits of our experimental resolution the data are consistent
with a single state, and having no evidence for two states we set upper
limits on the mass difference between two hypothetical states for different
assumed ratios of contributions to the observed peak. 
For equal contributions, the
$95\%$ confidence level upper limit on the mass difference is $3.6\:\mbox{MeV}/c^2$.
Under the single-state model the \X{} mass is measured to be 
$3871.61 \pm 0.16\:\mbox{(stat)} \pm 0.19\:\mbox{(syst)}\:\mbox{MeV}/c^2$,
which is the most precise determination to date.
\end{abstract}

\maketitle

%===============================================================================
% Introduction 
%===============================================================================
The discovery of the \X \cite{X3872:Belle,X3872:CDF} 
and many additional unexpected states \cite{Godfrey:2008nc} has revived
general interest in spectroscopy in the charmonium mass region. 
Initial attempts to explain the
\X as a conventional bound state of a $c$-quark and
an anti-$c$-quark have shortcomings \cite{Barnes:2003vb} which
triggered the development of unconventional explanations. 
Two popular models are a molecular
state composed of $D^0$ and $\bar{D}^{*0}$ mesons 
\cite{Tornqvist:2004qy,Swanson:2003tb}, and
a four-quark state \cite{Maiani:2004vq}.

In efforts to resolve the nature of the \X, several of its 
properties have been measured. The first determinations of its
mass \cite{X3872:Belle,X3872:CDF,Abazov:2004kp,Aubert:2004ns} 
resulted in values very close to the $D^0\bar{D}^{*0}$ mass
threshold. The observed width in these measurements was compatible with zero.
Studies of the \X production properties in $p\bar{p}$
collisions \cite{Abazov:2004kp,Bauer:2004bc} suggest that the 
production mechanisms are similar to those for the
$\psi(2S)$ charmonium state.
Several measurements constrained
the quantum numbers spin ($J$), parity ($P$) and charge-conjugation
parity ($C$) of the \X. These include evidence
for the decay modes $\X \rightarrow \Jpsi\gamma$,
$\Jpsi\omega$, and $\psi(2S)\gamma$ \cite{Abe:2005ix},
and a measurement of the mass distribution of the dipions 
from the $\X\rightarrow\Jpsi\pi^+\pi^-$ decay \cite{Abulencia:2005zc}. 
These measurements indicate an even $C$ parity.
A subsequent angular analysis constrained the quantum
numbers to only two possibilities, $J^{PC}=1^{++}$ or
$2^{-+}$ \cite{Abulencia:2006ma}.
A possible further decay mode of the \X was identified as
a peak near threshold in the $D^{0}\bar{D}^0\pi^0$ 
invariant mass spectrum \cite{Gokhroo:2006bt} with a mean
mass more than 3 MeV/$c^2$ above measurements in the $\Jpsi\pi^+\pi^-$ mode.
Despite efforts on both the
experimental and theoretical sides, the nature of the \X still
remains an unresolved puzzle.

A measurement of the \X mass with increased precision can provide 
crucial information for understanding its nature.
Under the hypothesis of a molecular state the mass of the \X
has to be lower than the sum of the $D^0$ and $\bar{D}^{*0}$ masses.
The four-quark state hypothesis predicts the existence of
two distinct particles that differ by the light-quark content
bound to the $c\bar{c}$ quarks. These two particles
should have slightly different masses, and the model of Maiani et al.
\cite{Maiani:2004vq} predicts a mass
difference at the level of $8\pm3$ MeV/$c^2$.
Recent measurements of the difference between the \X mass in 
$B^+ \rightarrow \X K^+$ and $B^0 \rightarrow \X K^0$ decays
\cite{Aubert:2008gu,Belle:2008te} disfavor this model under the hypothesis 
that one state is dominantly produced in $B^+$ decays and the other
one in $B^0$ decays.

In this Letter we report a study of the mass of the \X
resonance produced in $p\bar{p}$ collisions.
We consider the conjecture that the structure observed in our data is
composed of two different states with distinct masses; but failing 
to discern any evidence for this possibility we set an upper limit on the
mass difference between two hypothetical states. 
In light of this result we perform a precision measurement of the \X mass,
the main result of this paper.

%%%%%%%%%%%%%%%%%%%%%%%%%%%%%%%%%%%%%%%%%%%%%%%%%%%%%%%%%
% Detector 
%%%%%%%%%%%%%%%%%%%%%%%%%%%%%%%%%%%%%%%%%%%%%%%%%%%%%%%%%
The data were collected by the CDF II 
detector at the Fermilab Tevatron $p\bar{p}$ collider
between February 2002 and August
2007, and correspond to an integrated luminosity of
$2.4$ $\mathrm{fb^{-1}}$.  
The CDF II detector \cite{Acosta:2004yw} consists of a magnetic
spectrometer surrounded by electromagnetic and hadronic 
calorimeters and muon detectors.  
The tracking system is immersed in a 1.4 T axial magnetic field and is composed of a silicon 
microstrip detector \cite{Hill:2004qb} surrounded by an open-cell
drift chamber (COT) \cite{Affolder:2003ep}.
It extends out to a radius of 138 cm with up to
96 position measurements in the COT, and achieves a transverse momentum resolution of
$\sigma(p_T)/p_T \approx 0.15\,\%\,p_T/($GeV$/c)$.
We detect muons in planes of multiwire drift chambers and scintillators \cite{Ascoli:1987av}
in the pseudorapidity range $|\eta| \le 1.0$.
Events with $\Jpsi \rightarrow \mu^+ \mu^-$
decays are recorded using a dimuon trigger, which requires
two oppositely-charged COT tracks matched to muon chamber
track segments. The reconstructed invariant mass of a dimuon pair is
required to be between 2.7 and 4.0 GeV/$c^2$.

%===============================================================================
% Data Reconstruction  and Selection
%===============================================================================
To reconstruct \X candidates we first build \Jpsi candidates by combining pairs of
oppositely charged muon candidates with a transverse momentum, $p_T$, larger than 1.5 GeV/$c$.
The \X candidates are formed by combining \Jpsi candidates 
in the invariant mass range from 2.95 to 3.25 GeV/$c^2$ with pairs of oppositely
charged tracks, each with $p_T > 0.4$ GeV/$c$ and assigned the pion mass. 
We require that all four tracks have at least 10 COT and 2 silicon hits.
For the resulting \X candidates with $p_T > 3.5$ GeV/$c$, we perform a 
kinematic fit in which the tracks 
are constrained to originate from a common vertex and
the dimuon invariant mass is constrained to the world average \Jpsi
mass \cite{PDG2008}.
Candidates having a kinematic fit of good quality are selected
in a broad invariant mass range containing, in addition to \X candidates, 
also $\psi(2S)$ candidates that decay to the same final state.
The $\psi(2S)$ serves as a valuable control sample.

\begin{figure}[t]
  \centering
  \includegraphics[width=2.6in]{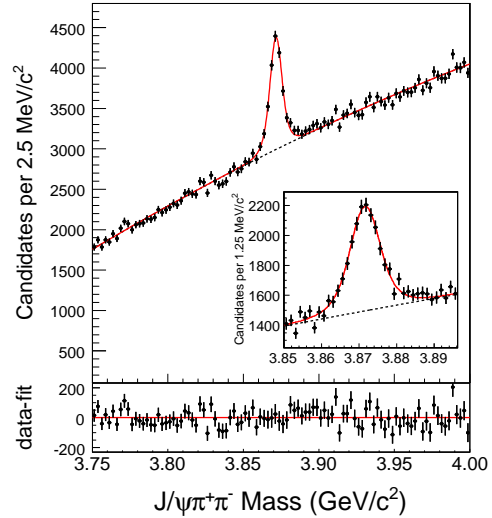}
  \caption{Invariant mass distribution of the \X candidates.
The points show the data distribution, the full line is the projection of the
unbinned maximum-likelihood fit, and the dashed line
corresponds to the background part of the fit.
The inset shows an enlargement of the region around the $X(3872)$ peak.
Residuals of the data with respect to the fit are displayed below the mass plot.}
  \label{fig:massDistribution}
\end{figure}

Several discriminating quantities are combined by a neural network 
into a single selection variable.
The individual quantities are transformed such that linear dependences
on the invariant mass are removed.
The most important inputs to the neural network are the $Q$ value
of the decay, defined as
$Q=m_{\Jpsi\pi^+\pi^-}-m_{\pi^+\pi^-}-m_{\Jpsi}$, the transverse
momenta of the two pions, the quality of the kinematic fit of
the \X candidate, and muon identification
quantities. The offline muon identification is based on 
the matching of tracks found in the tracking system to
track segments in the muon system and on the energy deposited
in the calorimeter by the muon-candidates.
For the training of the neural network, a background sample
is extracted from data, selecting events in regions of the 
$\Jpsi\pi^+\pi^-$ mass away from the \X and
$\psi(2S)$ signals, mainly consisting of \Jpsi particles combined with two random tracks. 
For the signal sample we use simulated
\X events. In the simulation we generate a single \X
per event using the momentum distribution of the $\psi(2S)$,
which is then decayed using the {\sc evtgen} package
\cite{Lange:2001uf}. Each event is then passed through
a detector simulation based on the {\sc geant3} package \cite{GEANT3} 
and a trigger simulation, and is reconstructed with
the same code as for real data. 
The simulation is in good agreement with the data
as verified with several kinematic quantities.
The final selection places a requirement on the
neural network output and the number of candidates per
event. Using wrong-sign candidates, where the two pion candidates
have the same charge, we verify that the selection
does not create an artificial excess in the mass spectrum.
The invariant mass distribution of the selected
candidates in the \X mass region is shown in
Fig.~\ref{fig:massDistribution}. 
The sample contains about 6000 \X signal events.
%%%%%%%%%%%%%%%%%%
% Fig. 1
%%%%%%%%%%%%%%%%%%
%%%%%%%%%%%%%%%%%%

Before we perform a mass measurement, we test whether the signal is
consistent with a single state
or we have evidence for more than one state.
In the test we perform a
binned maximum likelihood fit to the mass distribution in data, where we describe the combinatorial
background by a second order polynomial, and the signal by a
nonrelativistic Breit-Wigner function 
convolved with a resolution function determined from simulated events
and parametrized by the sum of two Gaussians. 
The core Gaussian, with a width of 3.2 MeV$/c^2$, accounts for two thirds 
of the resolution function; the second Gaussian has about twice the width.
In the fit we fix the width of the
Breit-Wigner function to $\Gamma= 1.34$ MeV$/c^2$, our average of the widths
measured in $\Jpsi\pi^+\pi^-$ decays \cite{X3872:Belle,Aubert:2008gu}.
The uncertainty on $\Gamma$ of $0.64$ MeV$/c^2$ is taken into account
in the hypothesis test described below.
As a test statistic we introduce a factor $t$
that scales the intrinsic and resolution widths of the signal shape.
The value of $t$ determined by the fit to the data is then
compared to the distribution of $t$ from
an ensemble of simulated experiments that assume a single state.
Based on this comparison the consistency of the data 
with the single-state hypothesis is evaluated.
The pseudoexperiments are generated using the same fit
model as in data. 
As several quantities are known only with limited
precision, we vary those in the sample generation according to their uncertainties.
The varied parameters include background shape parameters, the
number of signal and background events, the width of the
Breit-Wigner function, and the overall width of the resolution
function. 
From a comparison of
the $\psi(2S)$ signal in the data to that of simulated events we observe that
the simulation underestimates the resolution by about 5\,\%.
The samples were generated with a resolution corrected for this discrepancy.

\begin{figure}[t]
  \centering
  \includegraphics[width=2.6in]{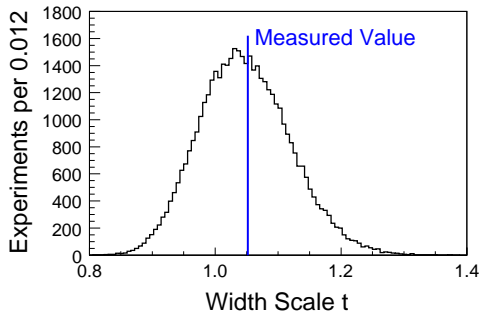}
  \caption{Distribution of the width scale $t$ for generated
experiments using the single state hypothesis (histogram).
Also shown is the measured value from data (vertical line).     }
  \label{fig:scaleNull}
\end{figure}

\begin{figure}[t]
  \centering
  \includegraphics[width=2.6in]{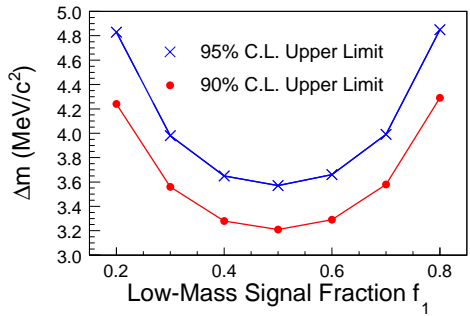}
  \caption{ The upper limit on the mass difference $\Delta m$  between
two states as a function of the fraction $f_1$ of the yield of the lower mass state. }
  \label{fig:limit}
\end{figure}

From data we obtain a width scale parameter value
of $t=1.052$. In Fig.~\ref{fig:scaleNull} we show a comparison of the fitted scale parameter to
the distribution obtained from simulated experiments assuming a single state. 
We conclude that the data 
are fully consistent with a single state.
In the absence of evidence for two distinct states
we set an upper limit on the possible mass difference
between two hypothetical states. As a test statistic we use
the width scale $t$, which is compared to expectations from
samples simulated with different mass splittings.  We assume
that both states have the same mass shape and do not interfere. We derive
upper limits as a function of the fraction $f_1$ of the lower
lying state to the total observed signal. The resulting $90\%$ and $95\%$ C.L.
upper limits are shown in Fig.~\ref{fig:limit}.
For an equal mixture of the two contributing states, the
limits are
$\Delta m<3.2$ MeV$/c^2$ and $\Delta m<3.6$ MeV$/c^2$ at $90\%$ and 
$95\%$ confidence levels, respectively.
This result is complementary to other measurements 
\cite{Aubert:2008gu,Belle:2008te} in that
it does not rely on assumptions about the production of the two
hypothetical states in $B^+$ versus $B^0$ decays, but
depends on $f_1$.
%%%%%%%%%%%%%%%%%%
% Fig. 2
%%%%%%%%%%%%%%%%%%
%%%%%%%%%%%%%%%%%%
%%%%%%%%%%%%%%%%%%
% Fig. 3
%%%%%%%%%%%%%%%%%%
%%%%%%%%%%%%%%%%%%

Lacking any indication of dual states we proceed to extract
the mass of the \X by performing an unbinned
maximum likelihood fit using the same fit model as used 
in the previous two-state test. In this fit we fix
the intrinsic width to $\Gamma=1.34$ MeV$/c^2$ and the resolution
parameters to their expected values. Free
parameters in the fit are the mass of the \X,
the fraction of signal events in the
sample, a resolution scale factor,
and two parameters determining the background shape.

To check the absolute mass scale we use the nearby $\psi(2S)$ signal 
in the same $\Jpsi\pi^+\pi^-$ invariant mass spectrum.
We use the identical fit model as for the $X(3872)$, with the exception
that the signal shape parameters are adjusted to the world
average value of $\Gamma=0.337$
MeV/$c^2$ \cite{PDG2008} for the intrinsic width, and
that resolution parameters are determined from simulated $\psi(2S)$ events.
The fit yields $m_{\psi(2S)}=3686.03\pm0.02\:\mbox{(stat)}\:\mbox{MeV}/c^2$.
While this value is consistent with the world average $\psi(2S)$ mass
of $3686.09\pm0.03\:\mbox{MeV}/c^2$ \cite{PDG2008},
we use the $60$ keV$/c^2$ difference
between our measurement and the world average value as an
estimate of a possible uncertainty due to uncertainties both
on our measurement and on the world average value.

Since a possible miscalibration of the momentum scale would
show up as a dependence of the measured mass on momentum, we
measure the $\psi(2S)$ mass as a function of several kinematic
variables.
We find that any tested dependence has an effect
below $0.1$ MeV/$c^2$, which is taken as an additional measure of
the systematic uncertainty.
This uncertainty is summed in quadrature with the systematic uncertainty
on the absolute mass scale derived above.
To translate the estimation of the
mass-scale uncertainty from the $\psi(2S)$ to the
$X(3872)$ we scale the sum by a factor of 1.6 that is modeled by a linear dependence
on the mass with respect to the $\Jpsi\pi^+\pi^-$ threshold.
This yields a total systematic uncertainty of
$0.19$ MeV/$c^2$ attributed to the momentum scale.

To estimate the effect due to the uncertainties in the fit model, we
refit the data using alternative models. These include the use of a linear
function instead of a second-order polynomial for the
background description, a single Gaussian function instead of
a non-relativistic Breit-Wigner function convolved with
double Gaussian resolution function for the signal
description, and fixing the natural width $\Gamma$ to zero or to twice the
nominal value. We also perform a fit in a mass window
reduced by 40\%. All of these modifications have a negligible effect
on the fitted mass, below 20 keV$/c^2$, and therefore we do not 
assign any systematic uncertainty to the measurement due to the fit model.
Because the observed decays to $D^0\bar{D}^{*0}$ may stem from a different particle
we assume that the mass line shape is not distorted by them.
If this were the case, as discussed in Ref.\ \cite{Braaten:2007dw},
it would be expected to increase the measured mass by about 150 keV$/c^2$.

The final mass measurement for the
$X(3872)$ is
$3871.61\pm0.16\:\mbox{(stat)}\pm0.19\:\mbox{(syst)}$ MeV/$c^2$.
The measured value is in good agreement with the world average \cite{PDG2008}
and the more precise average of measurements in the $\Jpsi\pi^+\pi^-$ channel
including the preliminary Belle measurement \cite{Belle:2008te}.
It is the most precise single measurement to date and
improves the precision of the latter average
by about a factor of 1.5.

%===============================================================================
% Discussion 
%===============================================================================
Our measurement is below the $D^0\bar{D}^{*0}$ mass threshold of $3871.80\pm0.35$
MeV/$c^2$ \cite{PDG2008} by $0.19\pm 0.43$ MeV/$c^2$. This implies that
the interpretation of the \X as $D^0\bar{D}^{*0}$ molecule is still
possible, although the current precision does not preclude
an \X mass above the $D^0\bar{D}^{*0}$ mass threshold.
A future increase in precision of this
comparison will therefore require improvements in the precision of the $D^0$ and $D^{*0}$ masses. 
Concerning the four-quark hypothesis, our mass splitting upper limits
for two hypothetical states with relative fractions between 0.2 and 0.8
exclude the range of $8\pm 3$ MeV/$c^2$ predicted in Ref.~\cite{Maiani:2004vq}.

%===============================================================================
% SUMMARY
%===============================================================================
In summary, we present a new measurement of the $X(3872)$ mass
using its decay to $\Jpsi\pi^+\pi^-$. Our measured value
of $3871.61 \pm 0.16\:\mbox{(stat)} \pm 0.19
\:\mbox{(syst)}\: \mbox{MeV}/c^2$
supersedes that of Ref.~\cite{X3872:CDF}, and
is more than two times more precise than the best single measurement so far.
In addition, we derive upper limits on the mass difference for the
hypothesis of two \X states, which are predicted by some four-quark
scenarios, as a function of their relative contribution to the observed signal. 
For an equal mixture of the two possible states, the limit is
$\Delta m<3.6$ MeV$/c^2$ at $95\%$ confidence level.

%===============================================================================
% ACKNOWLEDGEMENTS
%===============================================================================
We thank E.~Braaten for useful discussions.
We thank the Fermilab staff and the technical staffs of the participating 
institutions for their vital contributions. This work was supported by 
the U.S. Department of Energy and National Science Foundation; the 
Italian Istituto Nazionale di Fisica Nucleare; the Ministry of Education, 
Culture, Sports, Science and Technology of Japan; the Natural Sciences 
and Engineering Research Council of Canada; the National Science Council 
of the Republic of China; the Swiss National Science Foundation; the A.P. 
Sloan Foundation; the Bundesministerium f\"ur Bildung und Forschung, 
Germany; the Korean Science and Engineering Foundation and the Korean 
Research Foundation; the Science and Technology Facilities Council and the Royal 
Society, UK; the Institut National de Physique Nucleaire et 
Physique des Particules/CNRS; the Russian Foundation for Basic Research; the
Ministerio de Ciencia e Innovaci\'{o}n, and Programa Consolider-Ingenio 2010, Spain;
the Slovak R\&D Agency; and the Academy of Finland.

%===============================================================================
% BIBLIOGRAPHY
%===============================================================================

\end{document}